# Superresolution optical magnetic imaging and spectroscopy using individual electronic spins in diamond

Jean-Christophe Jaskula[1,2,3,*], Erik Bauch[1,3,*], Silvia Arroyo-Camejo[4], Mikhail D. Lukin[1], Stefan W. Hell[4], Alexei S. Trifonov[1,2,3,5], Ronald L. Walsworth[1,2,3,#]

[1] *Department of Physics, Harvard University, Cambridge, Massachusetts 02138, USA*
[2] *Harvard-Smithsonian Center for Astrophysics, Cambridge, Massachusetts 02138, USA*
[3] *Center for Brain Science, Harvard University, Cambridge, Massachusetts 02138, USA*
[4] *Department of NanoBiophotonics, Max Planck Society, Am Fassberg 11, 37077 Göttingen, Germany*
[5] *Ioffe Physical-Technical Institute RAS, Saint Petersburg, Russia*
[*] *These authors contributed equally to this work*
[#] *Corresponding author: rwalsworth@cfa.harvard.edu*

*Nitrogen vacancy (NV) color centers in diamond are a leading modality for both superresolution optical imaging and nanoscale magnetic field sensing. In this work, we solve the remaining key challenge of performing optical magnetic imaging and spectroscopy selectively on multiple NV centers that are located within a diffraction-limited field-of-view. We use spin-RESOLFT microscopy to enable precision nanoscale mapping of magnetic field patterns with resolution down to ~20 nm, while employing a low power optical depletion beam. Moreover, we use a shallow NV to demonstrate the detection of proton nuclear magnetic resonance (NMR) signals exterior to the diamond, with 50 nm lateral imaging resolution and without degrading the proton NMR linewidth.*

NV centers in diamond are now the leading modality for nanoscale magnetic sensing, with wide-ranging applications in both the physical and life sciences, including the use of single NV center probes for imaging of magnetic vortices [1] and spin waves [2] in condensed matter systems as well as single proton magnetic resonance imaging (MRI) [3]; and the use of ensembles of NV centers for wide-field magnetic field imaging of biological cells [4, 5] and geoscience samples [6]. Many envisioned applications of NV centers at the nanoscale, such as determining atomic arrangements in single biomolecules [3] or realizing selective strong coupling between individual spins [7] as a pathway to scalable quantum simulations [8], would benefit from a combination of superresolution imaging techniques with high sensitivity NV magnetometry. Recently, mapping the position of multiple NV centers has been improved beyond the diffraction limit by techniques using magnetic field gradients [9-11], which locally shift the NV center resonances but can deteriorate the sample to be probed. Alternatively, far-field optical superresolution techniques have the advantage of being versatile, simple to integrate into standard NV-diamond microscopes, require no special fabrication technique or magnetic field gradients, are compatible with a wide range of NV sensing techniques, and allow for fast switching between multiple NV centers. Coordinate-stochastic superresolution imaging methods, namely STochastic Optical Reconstruction Microscopy (STORM) and Photo Activated Localization Microscopy (PALM), readily offer high parallelization in sparse samples, but are prone to artefacts at high emitter densities and have been implemented until now only for a few NV centers per diffraction limited volume [12, 13]. On the other hand, coordinate-deterministic superresolution methods provide targeted probing of individual NV spins with nanometric resolution [14-16], which is well suited for the purpose of coherent nanoscale AC magnetometry, where each NV acts as a local phase-controlled magnetometer probe.

In this letter, we demonstrate the capability of spin-RESOLFT (REversible Saturable OpticaL Fluorescence Transitions), a coordinate-deterministic technique for combined far-field optical imaging and coherent spin manipulation, to map spatially varying magnetic fields at the nanoscale, including the NMR signal from external nuclear spins. Importantly, spin-RESOLFT does not require multi-wavelength excitation and high optical powers, as typically used with STimulated Emission Depletion (STED) [17] microscopy or Ground State Depletion (GSD) by metastable state pumping [18]. As shown below, we use spin-RESOLFT to optically resolve individual NV centers with a resolution of about 20 nm in the lateral (xy) directions, while exploiting the spin-state dependent optical properties (Fig. 1(a)) and long electronic spin coherence times of NV centers in bulk diamond for precision magnetic field sensing. Moreover, we show that the localization along the beam propagation (z) axis can be simultaneously improved to sub-nanometer precision via NV NMR measurements from proton spins in a sample external to the diamond.

We first demonstrated how spin-RESOLFT allows imaging of NV centers with subdiffraction resolution given by [14] $FWHM \approx \lambda/[2NA\left(1 + \Gamma\tau_{doughnut}\right)^{1/2}]$ in the ideal case. Here, NA = 1.45 is the numerical aperture of the objective, $\Gamma$ is the optical pump rate, and $\tau_{doughnut}$ is the duration for which the doughnut beam is applied during the spin-RESOLFT experimental sequence (see Fig. 1 (c)). Fig. 1(d) shows examples of one-dimensional scans of a single

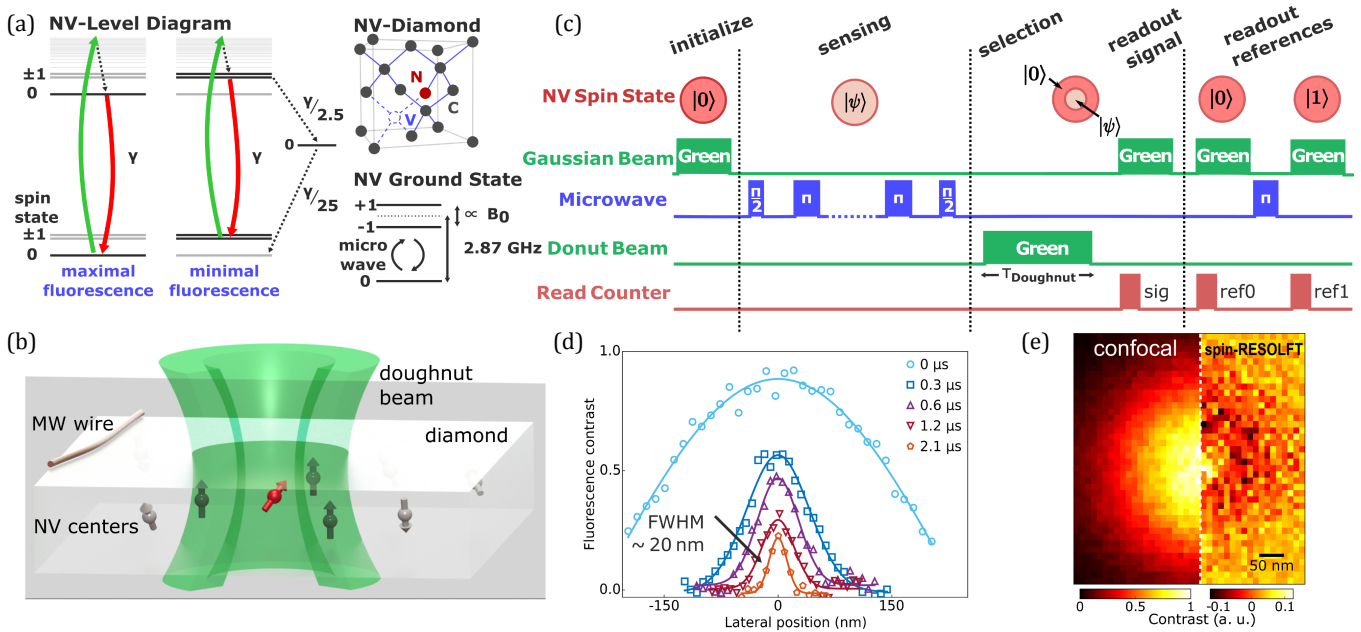

Fig. 1. Spin-RESOLFT imaging of NV centers. (a) Energy levels and diamond lattice schematic for the negatively charged NV center in diamond, which has electronic spin S=1. When electronically excited by green light absorption, the $m_s$=0 spin state largely fluoresces in the red with no change to the spin state, whereas the $m_S$ = 1 spin state has a significant probability to decay through a singlet state to the $m_s$=0 spin state, which effectively reduces the $m_s$=1 fluorescence rate and allows spin-state initialization into $m_s$=0. (b) The spin-RESOLFT experimental setup is merely an NV-diamond scanning confocal microscope augmented with a low power green doughnut beam. (c) Spin-RESOLFT experimental sequence for quantum sensing using NV centers in diamond, e.g., AC magnetometry for the spin echo pulse sequence shown. A spatially selective repolarization via the pulsed green doughnut beam is inserted before the spin readout to interrogate only a specific NV center. (d) 1D spin-RESOLFT scans for a single NV center and different doughnut durations. The superresolution intensity profiles are determined by comparing the fluorescence after applying the doughnut (pulse sig) with confocal scans (pulse ref0) (see Supplementary Information). (e) Similar resolution (≈35 nm) is achievable in a 2D spin-RESOLFT image.

NV center imaged after applying the doughnut beam at different durations. From numerical fits to our data using a five level model for the NV (see Supplementary Information), we extract FWHM = 20 ± 2 nm for a doughnut duration of 2.1 μs and a power of 700 μW, more than an order-of-magnitude improvement over confocal resolution. We note that the duration of the selective doughnut beam pulse (few microseconds) has minimal effect on the total sequence time (few hundred microseconds). Moreover by adjusting $\tau_{doughnut}$, sub-diffraction NV images can be attained with doughnut powers as low as 25 μW. Importantly, owing to the long lifetimes of the states harnessed for NV separation, the optical powers required for superresolution are several orders-of-magnitude lower than those required for STED [15, 18]. Similarly, Fig. 1(e) shows a comparison of two-dimensional images of a single NV center acquired both without (left) and with (right) the doughnut beam applied before readout. In practice, the maximum optical resolution is limited by a non-vanishing field intensity at the center of the doughnut mode due to beam shaping imperfections [17], aberrations induced by the sample, as well as thermal and vibrational instabilities of the apparatus (see Supplementary Information).

Spin-RESOLFT allows us to manipulate and address individual NV centers within a diffraction limited volume. For example, Fig. 2(a) shows a confocal image of two NV centers that are separated by less than the diffraction limit and can therefore not be resolved by means of confocal microscopy. In comparison, when using spin-RESOLFT microscopy (Fig. 2(b)), the individual NVs are clearly distinguished and their positions are localized within an uncertainty of 5 nm. To demonstrate selective coherent measurements of NV spins using spin-RESOLFT, we begin with measuring the Hahn-echo coherence time ($T_2$) for each NV individually (Fig. 2(c)) by applying a π/2 – π – π/2 MW pulse sequence, followed by spin selective readout using the exact positions for $NV_1$ and $NV_2$ extracted from Fig. 2(b). We find that although the two NVs are subject to a nominally similar spin bath in the diamond sample, the measured $T_2$ for each NV spin differs due to slight variations in the local environment. The NV ensemble spin coherence time measured in confocal mode is consistent with an average of the two individual NV $T_2$ values measured with spin-RESOLFT, weighted by the fluorescence collected from each single NV center. Indeed, due to a slight systematic mismatch between the doughnut and the Gaussian beam centers, the black square does not lie exactly in the middle of the crosses indicating the NV positions and therefore the NV ensemble $T_2$ (inset Fig. 2(c)) is correspondingly closer to the $T_2$ of $NV_2$ as measured with spin-RESOLFT.

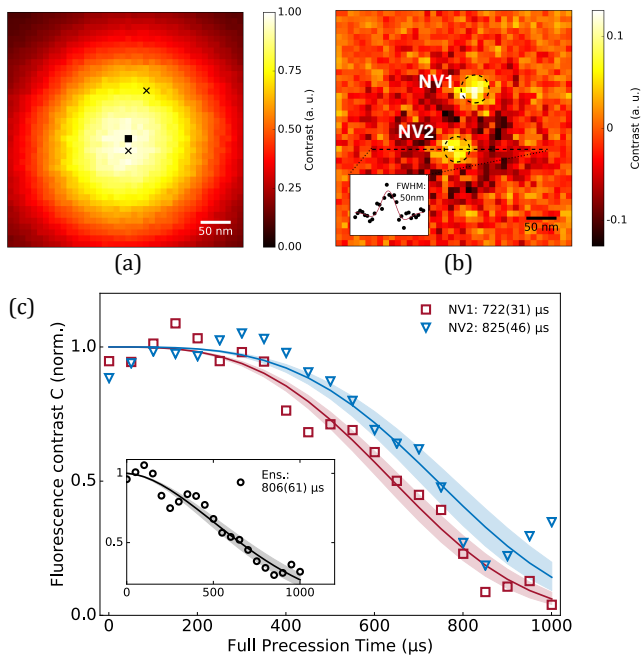

Fig. 2. Spin coherence time measurement for two NV defects resolvable only via spin-RESOLFT. (a) 2D Confocal image of two unresolved NV centers with the same orientation of their spin quantization axes. Black crosses indicate the NV positions as extracted from the spin-RESOLFT image, the black square indicates the Gaussian green laser beam center. (b) 2D spin-RESOLFT image of same field-of-view as in (a) (acquisition time 9 s per pixel). Inset: 1D intensity profile (dashed line) through NV2; the 50 nm FWHM is extracted using a numerical fit of a five-level model. (c) Spin coherence measurements and associated fits to a stretched exponential for the two NV centers shown in (a) and (b) Inset: coherence time determined for the ensemble of two NVs via a confocal measurement and associated fit.

Next, we demonstrated the utility of spin-RESOLFT to deploy each NV within a confocal volume as a very-well-localized, point-like quantum sensor. First, we selectively measured the response of $NV_1$ and $NV_2$ to an externally and spatially varying AC applied magnetic field. The field is produced by an AC current that runs through a wire at a 10-micron distance (Fig. 3(a)). The resulting magnetic field gradient, $\Delta B/\Delta r \approx 1$ nT/nm, leads to a measurable difference in field strength for $NV_1$ and $NV_2$. In Fig. 3 (b) we plot the measured coherence signal of $NV_1$ and $NV_2$ obtained for different magnetic field strengths by incrementally varying the magnitude of the AC current through the wire. The observed oscillations in NV fluorescence contrast are characteristic for spin-based local magnetometry [19] and their synchronization ensures that both sensors are identically calibrated. At a fixed current $I_{AC}$ = 7 mA (dashed lines, Fig. 3(b)), we measured a magnetic field of 8.924 ± 0.004 µT for $NV_1$ and 8.812 ± 0.009 µT for $NV_2$, which is in good agreement with the expected magnetic field profile of the wire (see Supplement). The field strength difference is better highlighted by plotting the Fourier transform of the coherence signals as function of the input current (Fig. 3(c)). As with the NV spin coherence time measurements (Fig. 2), the magnitude of the AC magnetic field found in confocal mode depends systematically on the position of the Gaussian beam and is only a weighted average of the magnetic field magnitudes determined individually for $NV_1$ and $NV_2$ using spin-RESOLFT (Fig. 3(d)).

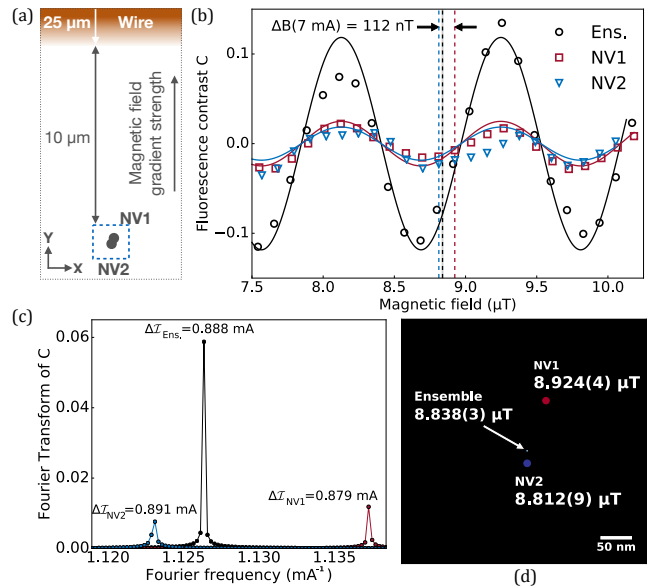

Fig. 3. Superresolution magnetic field imaging for two NV centers via spin-RESOLFT. (a) Schematic of the wire and two NV centers (same as in Fig. 2(b)). (b) spin-RESOLFT AC magnetometry measurements at 8.3 kHz for each NV center individually and for the two NV ensemble in confocal mode. Also shown are fits of data to sinusoids with phase fixed to zero for no applied current. Vertical dashed lines indicate the field strength for a peak current of 7 mA. (c) Fourier transforms of the measured response (Fig. 3(b)) of each NV center (blue and red linked dots) and the ensemble of two NV centers (black linked dots) to the external AC magnetic field. (d) 2D magnetic field map created by spin-RESOLFT and confocal measurements. The size of the discs for each NV is given by the fit uncertainty of the position from the superresolved NV imaging.

To show the applicability of NV spin-RESOLFT for nanoscale magnetic imaging and spectroscopy, we used a shallow NV center located approximately 3 nm below the diamond surface (see below), and simultaneously imaged the NV lateral position with sub-diffraction resolution of 50 nm and sensed the NMR signal from a statistically-polarized nanoscale sample of protons in immersion oil on the diamond surface. Shallow implanted NV centers are a promising modality for quantum computing [8], nanoscale magnetic resonance imaging [20] and single molecule detection [3] due to the strong dipolar and hyperfine interactions with electronic [10] and nuclear [21, 22] spin species located on the diamond surface. Adversely, surface effects tend to shorten the Hahn-echo $T_2$ of shallow NVs [23, 24], typically to tens of microseconds, which consequently leads to a reduction in magnetic field sensitivity. Thus we integrated spin-RESOLFT with an XY dynamic decoupling protocol to extend [25] the shallow NV $T_2$ and enable practical nanoscale NMR imaging (Fig. 4(a)). The dynamic decoupling protocol creates a coherent superposition of the NV $m_s = 0$ and $m_s = 1$ spin states, and then alternates this spin coherence between free evolution (of duration τ) and π phase flips, before converting the total accumulated phase into an NV spin state population that is measured optically (Fig. 4(b)). We find that spin-RESOLFT can be combined with dynamical decoupling sequences to increase the NV coherence time up to 100 μs while providing superresolution. Moreover, the NV spin phase accumulation is strongly perturbed when a frequency component of the external magnetic field matches twice the free evolution period $\tau = \nu_B/2$. Thus we observed a spectrally narrow dip in the NV coherence signal (Fig. 4(c)) at the proton spin Larmor precession frequency $\upsilon_p = (\gamma_p/2\pi)B_0 \approx 1.2$ MHz, where $\gamma_p$ is the proton spin gyromagnetic ratio and $B_0$=282 G is the applied static magnetic field, which is indicative of a NMR signal from statistically-polarized proton spins in the immersion oil on the diamond surface [21]. We also find that applying a 10 μs long doughnut beam pulse of 30 μW average power does not deteriorate the proton NMR signal, while allowing for far-field optical spin readout of a sub-diffraction sized area with a lateral diameter of around 50 ± 5 nm. Furthermore, by fitting the NV NMR data to an analytical model (see supplementary material), we determined the depth of the NV quantum sensor below the diamond surface to be 3.0 ± 0.3 nm [26].

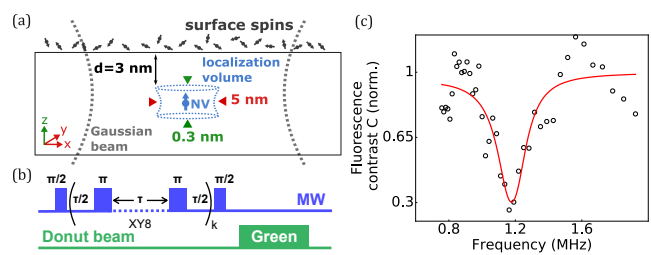

Fig. 4. NV spin-RESOLFT sensing of proton NMR. (a) Schematic showing nanometer-scale localization volume of a shallow NV. (b) XY8-k dynamical decoupling pulse sequence used for NMR proton sensing with sub-diffraction resolution. (c) NV spin-RESOLFT NMR spectroscopy of proton spins in immersion oil on the diamond surface (black dots) and fit to an analytical model [26] (red curve).

Our results demonstrate five key features of the spin-RESOLFT technique when applied to NV centers in diamond: (i) selective NV imaging and coherent spin manipulation for multiple NVs within a confocal volume with no corrupting effect on NV spin coherence; (ii) sensitive nanoscale magnetic imaging and spectroscopy via a technically straightforward, far-field optical technique; (iii) compatibility with dynamical decoupling sequences to extend NV $T_2$; (iv) operation with much lower optical depletion power than conventional deterministic superresolution imaging methods; and (v) applicability to shallow NVs for applications such as nano-NMR. For an imaging resolution of 20 nm, as demonstrated here, the spin-spin interaction between individual NV centers (~10 kHz) is larger than their typical decoherence rate (~1 kHz), fulfilling a fundamental requirement for quantum information applications [7, 27]. Furthermore, low-power superresolution imaging techniques can be critical for many applications, such as those that require cryogenic temperatures or shallow NV centers or for light-sensitive biological samples, as high optical power may cause heating as well as surface and sample deterioration. We also expect that the spin-RESOLFT technique can be straightforwardly extended to other NV-based sensing modalities, including temperature [28], electric field [29], and charge state [30] detection with nanoscale optical resolution.

*Acknowledgments*. This work was supported by DARPA (QuASAR program), DOD MURI quantum science programs, the NSF, the National Security Science and Engineering Faculty Fellowship program, and the Gordon and Betty Moore Foundation.

## Supplementary Material

### Diamond Sample Information

Sample A used in Fig. 1(d), Fig. 1(e), Fig. 2 and Fig. 3 of the main text is an ultra-pure CVD diamond, isotopically engineered (99.99% $^{12}C$) with NV orientation along two of four crystal axes, spin coherence time ($T_2$) approaching one millisecond, and spin lattice relaxation time $T_1$ of a few milliseconds at room temperature. Sample B used for measurements shown in Fig. 4 is also an ultra-pure CVD sample, isotopically engineered (99.999% $^{12}C$) with shallow implanted NV centers 1 - 20 nm below the surface ($^{14}N$ at 2.5 keV). Both samples were created by Element Six.

### Spin-RESOLFT Imaging

By first applying a π-pulse to switch all the NV centers into the spin state $m_s=1$ and then using a selectively repolarizing green doughnut beam, we pump off-center NVs into the $m_s=0$ ground-state. These off-center NV centers contribute a spatially broad 'background' fluorescence signal in addition to a spatially narrower fluorescence feature characteristic of superresolved NV centers in the center of the doughnut beam (green curve in Fig. S1(a). We determine the background from the off-center NVs by recording a confocal scan (blue curve in Fig. S1(a) immediately following the scan acquired with the doughnut beam. By subtracting the two signals, we obtain the 1D spin-RESOLFT image, which displays a non-Gaussian intensity profile (Fig. S1(b). The observed profile is strongly dependent on the degree of NV spin repolarization that occurs when the doughnut beam is applied, which is discussed in detail in the next section. We note that the intensity profiles in Figure 1(d) of the main text were taken under conditions of short doughnut pulse duration, permitting us to approximate the linewidth as Gaussian.

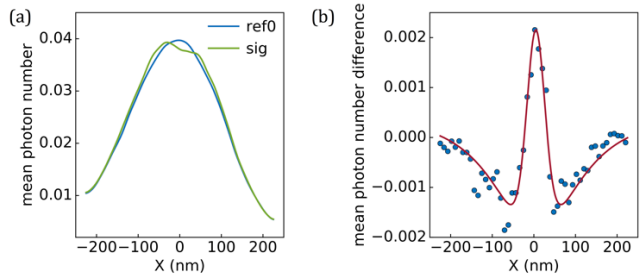

Fig. S1 (a) Single NV fluorescence measurements as a function of relative position (1D) acquired for the spin-RESOLFT protocol: after the application of the doughnut beam (signal, green) and after a complete repolarization with a Gaussian beam (ref0, blue). A 2-pixel running average is applied to smooth shot-noise-limited intensity fluctuations. At certain positions, the NV spin repolarization occurring from doughnut beam illumination is more efficient, leading eventually to a stronger fluorescence signal. (b) 1D spin-RESOLFT NV image (blue dots) constructed by subtracting the fluorescence curves shown in **a**. Red curve is a numerical fit of data to a five level model (see next section).

In Figure2(b) of the main text, we show that two NV centers within the same diffraction-limited volume are distinguished by the spin-RESOLFT technique. From the correlated spin-RESOLFT image we extract the distance between the two NV centers to be d = 105 ± 16 nm. Due to a ≈ 20 nm misalignment between the Gaussian readout and doughnut beams, the maximum fluorescence in the confocal image is not perfectly aligned with the axis formed by the two NV centers.

### NV Spin Coherence Measurements

There is a rapid dephasing of freely precessing NV spins on a time scale $T_2^* \sim$ 0.1-10 μs for typical spin impurity concentrations in diamond. By applying a single resonant MW pulse to refocus the dephasing, the Hahn-echo sequence decouples NV spins from spin bath magnetic field fluctuations that are slow compared to the free precession time. In diamond Sample A used in Fig. 2 and Fig. 3, the low impurity concentration leads to a long NV Hahn-echo spin coherence time $T_2$ of about 800 μs. Such coherence times are extracted from fits of Hahn-echo measurements of NV spin coherence to a stretched exponential coherence function $C(t) = A \exp\left(-\frac{t}{T_2}\right)^p$, where the parameter p is related to the spin bath surrounding the NV center [1]. For spin-RESOLFT measurements of individual NVs, $p_{NV1} = 3.2 \pm 0.3$ and $p_{NV2} = 3.5 \pm 0.5$ are found, which are in good agreement with the expected value p = 3 for a spin bath with Lorentzian spectral density [1, 2]. In confocal mode, the incoherent dynamics of these two NV centers results in a reduced value for $p = 1.7 \pm 0.2$ as expected.

### Proton NMR Measurements

XY8-k pulse sequences are applied to a single shallow NV to measure the NMR signal produced by ~100 statistically-polarized protons spins in immersion oil placed on the diamond surface. These pulse sequences produce NV spin phase accumulation that is transferred to a spin state population difference by means of the last microwave π/2 pulse. The choice of the phase of this last pulse allows for projections onto each NV spin state $m_s=0$ and $m_s=1$, resulting in fluorescence measurements $F_0$ and $F_1$. Common-mode noise from laser fluctuations is suppressed by normalizing the fluorescence signals together in a fluorescence contrast C = ($F_0$ - $F_1$) / ($F_0$ + $F_1$). NV sensing of the magnetic field Fourier components at frequencies υ is realized by measuring the fluorescence contrast C over a range of free evolution times $\tau = \upsilon/2$. NV spin "background" decoherence is characterized by slow exponential decay of the fluorescence signal over hundreds of microseconds (Fig. 2(c) of the main text). This background decoherence is fit to a stretched exponential function and normalized out, leaving only the narrower proton-NMR-induced dip in NV signal contrast on top of a flat baseline, as shown in Fig. 4(b). The shape of this dip is determined by the magnetic field fluctuations produced by the dense ensemble of proton spins in the immersion oil on the diamond surface, as well as by the filter function corresponding to the XY8-k dynamical decoupling pulse sequence. The magnetic field signal has cubic dependence on the distance between the NV center and diamond surface ($B_{RMS} \propto d_{NV}^{-3}$), which can be then estimated by fitting the dip with the following formula: $C(\tau) \approx \exp\left(-\frac{2}{\pi^2}\gamma_e^2 B_{RMS}^2 K(N\tau)\right)$. Here $B_{RMS}$ is the RMS magnetic field signal produced at the Larmor frequency by the proton spins, $K(N\tau)$ is a functional that depends on the pulse sequence and the nuclear spin coherence time, and N is the number of pulses, which are separated by the NV spin free precession time τ. A thorough derivation of this formula as well as the description of the functional $K(N\tau)$ is presented by Pham et al. [3].

## NV Spin Repolarization

Due to the spin dependent intersystem crossing through its singlet states, NV centers preferentially decay into the $m_s=0$ ground state under green illumination. This results in strong spin polarization after several excitation cycles. In spin-RESOLFT, the role of spin polarization is two-fold: preparing the initial NV state in $m_s=0$ for sensing using a Gaussian beam, and repolarizing the NV center into $m_s=0$ for superresolution imaging using the doughnut beam. As mentioned in the main text and shown in the inset of Figure 2 and Fig. S1, the fluorescence point-spread function (PSF) displays a non-trivial shape that is the result of non-linear NV repolarization. To model the NV polarization dynamics, we use the 5-level system shown in Fig. S2.

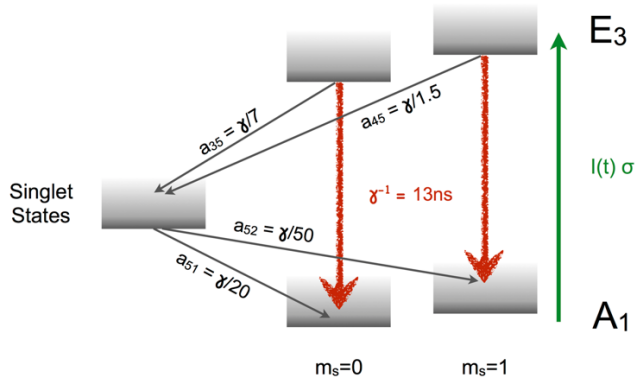

Fig. S2 NV level structure and decay rates. The populations are denoted by $n_i$, where i refers to the following levels: 1 for $m_s=0$ ground state, 2 for $m_s=-1$ ground state, 3 for $m_s=0$ excited state, 4 for $m_s=-1$ excited state and 5 for the singlet states. The decay rates $a_{ij}$ between levels are indexed by the initial level *i* and the final level *j*. All rates are given relative to the primary fluorescence decay rate γ. The singlet states are represented as a single state for the sake of simplicity, and we use previously measured room temperature rates [4].

The system of rate equations that governs the NV state populations under optical excitation can be formulated as

$$\frac{1}{\gamma}\frac{dn_1}{dt} = -I(t)\sigma \cdot n_1 + n_1 + a_{51}n_5 \quad (S1)$$

$$\frac{1}{\gamma}\frac{dn_2}{dt} = -I(t)\sigma \cdot n_2 + n_2 + a_{52}n_5 \quad (S2)$$

$$\frac{1}{\gamma}\frac{dn_3}{dt} = I(t)\sigma \cdot n_1 - n_3 - a_{35}n_3 \quad (S3)$$

$$\frac{1}{\gamma}\frac{dn_4}{dt} = I(t)\sigma \cdot n_2 - n_4 - a_{45}n_4 \quad (S4)$$

$$\frac{1}{\gamma}\frac{dn_5}{dt} = a_{35}n_3 + a_{45}n_4 - a_{51}n_5 - a_{52}n_5 \quad (S5)$$

Here σ represents the cross-section of the primary NV electronic transition for a 532 nm laser beam pulse of intensity I(t). Fig. S3 shows the $m_s = 0$ ground state population after applying a square pulse starting at t = 0 on an unpolarized NV center with equal initial spin state population. We see that the degree of repolarization depends on both the intensity and duration of the excitation pulse. In particular, a higher degree of polarization is achieved with a long and weak green pulse (5 ms at 5% of the saturation intensity). Moreover, for a fixed pulse duration, we find that the repolarization is non-linear, resulting in a strong effect on the PSF of the spin-RESOLFT microscope image determined by the spatial intensity distribution of the doughnut beam.

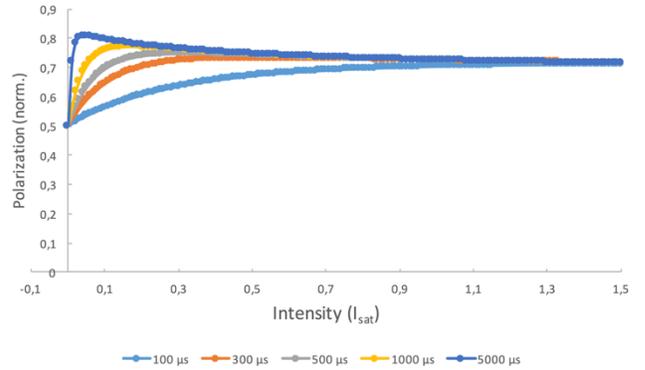

Fig. S3 Simulation of the dependence of the NV spin polarization on green excitation beam intensity. The degree of polarization displays a non-linear behaviour with light excitation. Short pulses of about 100 μs provide at most 70% polarization. Longer pulses provide a higher degree of polarization but restrict the intensity to a fraction of the saturation intensity. The highest resolution is obtained for durations where the slope near the doughnut center is steeper, which leads to strong non-linear behaviour, a degradation of the spin polarization far from the doughnut center and a non-trivial PSF profile.

Indeed, the doughnut intensity profile can be approximated near the center as

$$I(r) = I_0\left(\left(\frac{r}{r_0}\right)^2 + \epsilon\right)\exp\text{-}\left(\frac{r}{r_0}\right)^2 \quad (S6)$$

where $I_0$ is the peak intensity, $r_0$ is the doughnut radius, and $\epsilon$ is the relative residual intensity in the doughnut center. Using this intensity profile as input to the system of eq. (S1 – S5), we plot the one-dimensional spin-RESOLFT PSF in Fig. S4 for two different values of $\epsilon$ = 0.1 % and $\epsilon$ = 2 %. As the doughnut intensity increases quadratically, the degree of NV polarization varies according to the behaviour displayed in Fig. S3. Higher resolution is achieved for combinations of long durations and weak powers, which display strong repolarization. However, for a particular finite position in the doughnut profile, the intensity reaches the value where the repolarization is maximum, which leads to non-Gaussian wings in the spin-RESOLFT PSF. Doughnut imperfections, which lead to a non-zero intensity $\epsilon I_0$ in the doughnut beam center, tend to reduce the state dependent fluorescence contrast, but do not affect the shape of the intensity profile. We use the numerical solution to eq. (S1 – S5) to extract the resolutions reported in Figure 1(d), 1(e) and 2(b) in the main text.

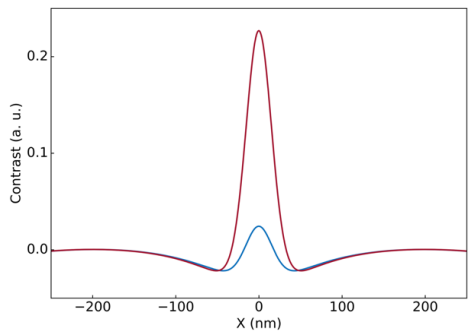

Fig. S4 Simulated spin-RESOLFT PSF for two different residual intensities in the center of the green doughnut beam: $\epsilon$ = 0.1 % (red) and $\epsilon$ = 2 % (blue).

## NV Position Drift and Fluctuations

Because of separated optical paths, the spin-RESOLFT experimental setup is sensitive to the relative motion of the Gaussian beam, doughnut beam, and confocal PSF, over the typical timescale of a complete experiment (minutes to hours). In particular, a single realization of a spin-RESOLFT experimental sequence requires ~20 µs, yielding ~0.02 collected photons. The sequence is repeated ~20,000 times for each imaging pixel to suppress photon shot noise to 5%. Thus a full 1D scan of ~400 nm (100 pixels) across an NV center ideally takes ~40 s. However, due to overhead from data recording and display, such a single 1D scan actually requires ~2 min. In addition, between each scan the position of the NV center is recorded, and then the optical illumination is adjusted to place the NV back into the middle of the scan window. The tracking procedure consists of discrete probing of the fluorescence spatial distribution in the neighbourhood of the NV center to determine the position of its maximum value. It is followed by 1D confocal scans in both lateral directions that are fitted with Gaussians to obtain the NV center position with a precision of about 5 nm. For the single NV spin-RESOLFT datasets plotted in Figure 1(d) of the main text, the entire 1D scan is repeated and then averaged 6 times, leading to a total acquisition time of about 12 min. In the case of multiple NV imaging (Figures 2 and 3 of main text), the tracking is done by taking a single nearby NV as reference. The reference NV is positioned about 1 µm away from the pair of NV centers, as shown in Fig. S5. In the more general case of a wide field-of-view image, optical reflection from a golden nanoparticle attached on the surface of the diamond is used as a reference point.

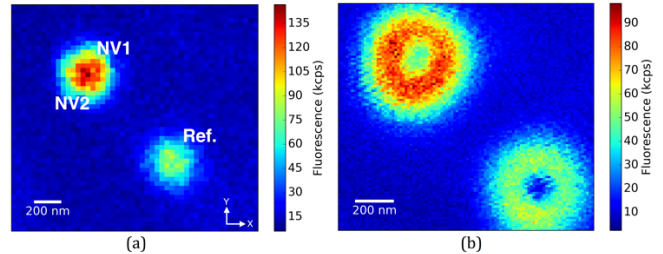

Fig. S5 Measured 2D images of pair of proximal NV centers and reference NV center (a) confocal scan; (b) green doughnut beam scan.

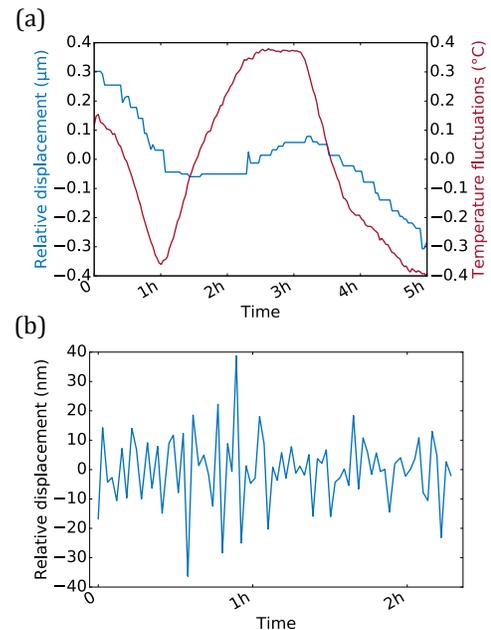

Fig. S6 (a) Measured relative 1D position of an example NV center and the laboratory temperature during a 5h-long confocal scan. A 1D NV fluorescence intensity profile takes about 1 minute after which the temperature and the NV center position are recorded. The laboratory temperature oscillates with a period of about 1 h and induces a correlated drift of the NV center position ~500 nm. (b) Stabilization of the laboratory temperature to a peak-to-peak variation of 0.1 °C allows data acquisition for two hours during which time the NV position is stable with a standard deviation of 11 nm.

During long acquisition times, the position of an NV center shows a strong correlation with laboratory temperature fluctuations, as shown in Fig. S6. Due to thermal expansion of the objective holder, we observed drifts of the reference NV by approximately 500 nm. These drifts were minimized by using insulating enclosures in which the temperature fluctuates by not more than 0.1°C over the course of a measurement. Nonetheless, such diminished drifts as well as table vibrations during a line scan can still result in observable broadening of the PSF of the spin-RESOLFT microscope. 2D scans, which are usually acquired over 10 hours, are affected even more severely. Fig. S6(b) displays the relative displacement of the NV center used in Figure 1(d) of the main text after each line scan. From this trace, we identify a motion along the direction of the scan with a standard deviation of 11 nm.

## AC Magnetic Field Gradient

To create an AC magnetic field gradient, which results in a measurable difference in magnetic field strength at the position of $NV_1$ and $NV_2$ as used for the results in Fig. 3 of the main text, we drive an AC current $I_{AC}$ = 7 mA at 8.3 kHz through a copper wire (type Alfa Aesar, diameter 25 μm) that is ~ 10 μm from the NVs. The same wire also carries the microwaves for coherent NV spin manipulation.

To simulate the observed magnetic field dependence, we devised a simple model that takes the projection of the applied AC fields onto the NV axis into account. In our geometry, the wire is parallel to the horizontal axis of Figures 2(b) and 3(d) (here, the y-direction), whereas the z-direction corresponds to the normal of the diamond's top surface and the x-direction completes the orthonormal reference frame. The NV center axis is determined by its polar θ and azithumal ϕ angles, as commonly defined. In this system of coordinates, $NV_1$ and $NV_2$ are directed along the x-direction (ϕ = 0°) while making an angle with the z-axis of θ = 54.7°. Moreover, the magnetic field lines form loops in the plane perpendicular to the wire. In AC magnetometry, the NV center is sensitive to the component of the magnetic field that is parallel to the NV axis, namely

$$B_\parallel (\vec{r}) = B_{wire}(\vec{r}) \cdot NV(\vec{r})$$
$$= \frac{\mu_0}{2\pi} \frac{I}{x^2 + z^2} (z \sin\phi \cos\theta + x \cos\phi). \quad (S7)$$

In Fig. S7(a), we simulate this magnetic field component for different positions along the x direction, with the center of the wire fixed at the origin and the NV center's depth is chosen to be at z = 12.5 + 7.5 μm = 20 μm under the wire. When the NV center is at x = -10 μm from the wire's edge, we find the calculated field along the NV center axis to be 9 μT, which is in good agreement with the experimental values we measure and report in the main text. Moreover, we plot the magnetic field gradient expected from this model as function of the x position in Fig. S7(b). The value of the magnetic field gradient of about 1 nT/nm is also in good agreement with the gradient experimentally measured with the pair of NV centers ($NV_1$ and $NV_2$) separated by 105 nm.

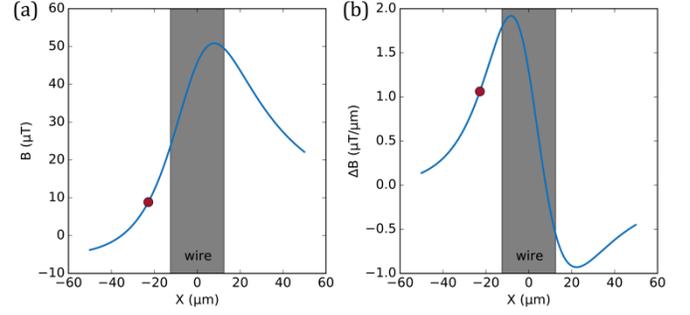

Fig. S7 (a) Magnetic field strength and (b) gradient extracted from a model that takes into account the distance from the wire and the orientation of the NV center axis. At a horizontal distance of 10 μm from the wire, the measured magnetic field strength of 9 μT and (b) measured gradient of 1 nT/nm are in good agreement with the experimental values reported in the main text.

The NV Rabi frequency's spatial dependence can also be derived from this model by considering the MW field component perpendicular to the NV axis, which is

$$B_\perp (\vec{r}) = B_{wire}(\vec{r}) - B_{wire}(\vec{r}) \cdot NV(\vec{r}). \quad (S8)$$

Using the same experimental conditions as described above, we calculate a Rabi frequency of 5.5 MHz for a current of 30 mA. This is also in a good agreement with the measured Rabi frequency for this MW current in our setup (Figure S8).

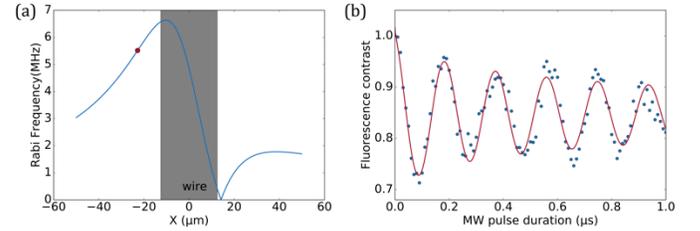

Fig. S8 (a) Calculated NV Rabi frequency as a function of the NV center horizontal position. This spatial behaviour is calculated from a model that takes into account the distance between the wire and the NV center as well as the NV orientation. The red dot corresponds to the position of the two NV centers used in the main text (NV1 and NV2). (b) Measured NV Rabi oscillations (blue dots) and a fit to an exponentially damped sinusoid (red curve). The extracted Rabi frequency of 5.5 MHz is a good agreement with the model calculation.